# Enriching gender in PER: A binary past and a complex future

Adrienne L. Traxler[1], Ximena C. Cid[2], Jennifer Blue[3], Ramón Barthelemy[4]

[1] Department of Physics, Wright State University, Dayton, OH 45435

[2] Department of Physics, University of Washington, Seattle, WA 98195

[3] Department of Physics, Miami University, Oxford, OH 45056

[4] Department of Teacher Education, University of Jyväskylä, Jyväskylä, Finland

ABSTRACT: In this article, we draw on previous reports from physics, science education, and women's studies to propose a more nuanced treatment of gender in physics education research (PER). A growing body of PER examines gender differences in participation, performance, and attitudes toward physics. We have three critiques of this work: (1) it does not question whether the achievements of men are the most appropriate standard, (2) individual experiences and student identities are undervalued, and (3) the binary model of gender is not questioned. Driven by these critiques, we propose a conception of gender that is more up-to-date with other fields and discuss gender-as-performance as an extended example. We also discuss work on the intersection of identities [e.g., gender with race and ethnicity, socioeconomic status, lesbian, gay, bisexual, and transgender (LGBT) status], much of which has been conducted outside of physics. Within PER, some studies examine the intersection of gender and race, and identify the lack of a single identity as a key challenge of "belonging" in physics. Acknowledging this complexity enables us to further critique what we term a binary gender deficit model. This framework, which is implicit in much of the gender-based PER, casts gender as a fixed binary trait and suggests that women are deficient in characteristics necessary to succeed. Alternative models of gender allow a greater range and fluidity of gender identities, and highlight deficiencies in data that exclude women's experiences. We suggest new investigations that diverge from this expanded gender framework in PER.

PACS: 01.40.Fk, 01.75.+m

## I. INTRODUCTION

Over the past several decades, physics education researchers have systematically studied the teaching and learning of physics. These efforts have focused extensively on students' conceptual understanding, development of curriculum to improve that understanding, cognitive aspects of learning, and preparation of future physics teachers [1,2]. In recent years, increasing attention has spread to a range of issues including student attitudes and epistemologies, affective factors, a sense of science or physics identity, and the learning communities in which students are situated [3–5].

As research areas broaden, the question of whether research findings apply equally to all students becomes increasingly salient. The underrepresentation of women, as well as African American, Hispanic, and Native American students, in physics is well documented,[1] and is an obvious area of concern for researchers invested in improving physics education. Barriers to women's participation in the field are widespread and range from minor to systemic in scope [6]. Additional theoretical perspectives are needed to address these broad challenges and their implications for the question of who benefits from education research. However, the conceptual frameworks used to treat gender in physics education research (PER) have remained largely unchanged over the past two decades. This relative stagnation is a marked contrast with other areas of PER, as noted below. We suggest it may be one contributing factor to the observation that dramatic widespread gains in the participation and success of women in physics classes have not materialized, despite evidence of such growth in conceptual gains [7].

Previous work on gender in PER has primarily incorporated gender as a fixed, binary, explanatory trait that may influence student conceptual or attitudinal gains, response to new curricula, and classroom success and retention. This work (reviewed in Section II) has provided a valuable foundation, but it must be expanded. PER has experienced such periods of growth in the past; for example, studies of student conceptual understanding provide one analogy. A great deal of early work in this area probed student conceptual knowledge, as well as gains in that knowledge, from a misconceptions-based framework. These investigations elaborated initial incorrect knowledge states and devised strategies for addressing students' deficiencies in content knowledge. However valuable this work might be, other theoretical frameworks for student learning opened many important new avenues for research. Aiming other lenses toward the construct of "conceptual understanding" brought tremendous advances in explanatory power, from phenomenological primitives or "knowledge in pieces" [8] to resources [9] to Vygotskian approaches that incorporate students' social environments [10].

Fortunately, an expanded framework for studying gender does not need to be built from scratch. Although it is not typically referenced by the PER community, this work has been ongoing in the fields of women's studies and gender studies for decades [11]. Our goal in this paper is to bridge this communication gap and suggest new avenues for physics education researchers who are interested in understanding the interplay of gender with physics education. An expanded view of gender inevitably tangles with other aspects of identity; therefore, we explore the first branches of these areas as well.

Section II of the paper surveys past and recent physics education research that incorporates gender. Section III introduces a different way of conceptualizing gender, as a performance rather than a fixed binary trait, and explores some consequences of this theoretical shift. Using these insights, Section IV discusses the intersection of gender with other facets of a person's identity. Section V suggests potential research directions that are available in an updated framework, and Section VI offers concluding thoughts on enriching gender in PER.

---

[1] See for example data from the American Physical Society (http://www.aps.org/programs/education/statistics/), sourced from the IPEDS Completion Survey.

*Glossary note:* Before proceeding further, a brief note on terminology is important.[2] In much of the literature reviewed, the terms *sex* and *gender* are used interchangeably. For reasons explored in the sections below, it is fundamentally important to separate these concepts.

- *Sex* refers to the biological and physiological characteristics of an individual. It is typically assigned at birth as male, female, or intersex [12].
- *Gender* refers to a perceived identity that may or may not align with biological sex. Gender is often used to describe an outward expression of clothing, accessories, outward appearance, or behaviors to signify masculinity and femininity that are validated (or not) by other members of society. Gender can also refer to an individual's internal perception of their identity, which may not be outwardly expressed. Neither of these facets of gender is necessarily fixed for an individual, so either or both may shift over time.
- *Transgender* (also trans or trans*) individuals have a gender identity that differs from their biological sex. They may or may not express this gender identity outwardly, depending on a range of factors including personal preference, social pressures, or workplace/classroom pressures.

Sex and gender are often not clearly distinguished in the literature. Some of the authors we cite below discuss sex differences, and others gender differences. This confusion of terms reinforces the implicit assumption that gender is fixed by biological sex. Sex itself is often cast as a strict binary, excluding intersex individuals; detailed discussion of this issue is outside the scope of this paper (but see [12–14] for an entry to the topic). In this paper, we will use "gender difference" and refer to the different genders as men and women. In some cases, that means that we are using different words than the authors we cite.

## II. RESEARCH ON GENDER IN PER

The first paper addressed at physicists that included research results regarding gender differences in the physics classroom was published in 1992, the year that many students starting graduate studies today were born. This paper, published in the American Journal of Physics, noted that problem-solving groups composed of two women and one man outperformed problem-solving groups one woman and two men [15]. These authors reported that this result held even when the one woman in the group was "articulate and the highest-ability student in the group" (page 641) because the men in the group might simply ignore her correct arguments.

---

[2] Here we borrow heavily from "Supporting LGBT+ Physicists & Astronomers: Best Practices for Academic Departments," available online at http://lgbtphysicists.org/files/BestPracticesGuide.pdf.

Just as the experiences of women students are sometimes ignored in their classes, they are often ignored as research subjects in PER (we return to this point in II.D below). That said, gender differences have been included in the PER literature in several topics, among them performance on standardized measures; preparation, interest, and retention; and the effects of reformed pedagogy.

Some research has also attempted to isolate gender difference as a factor by triangulating from a number of measures. Blue and Heller studied matched samples of men and women in an introductory university physics course. They found that when men and women were matched on a series of measures (three high-school variables; three pretests, including the FCI; their year in college; and locus of control), there was no difference in their performance on post-tests ( [16]; for more detail, see [17]). A comparable result was reported several years later at a different university [18], where most of the differences in performance at the end of a physics course were explained by differences in knowledge, attitudes, and beliefs at the start of the course. Further study showed that the differences between the genders at the end of the second physics course were primarily explained by differences at the end of the first physics course [19]. These accumulated differences might explain large differences in participation over time.

### A. Performance on Standardized Measures

A large body of influential work in PER concerns the development and implementation of standardized measures of student conceptual knowledge or attitudes. Many of these diagnostics, including the Force Concept Inventory (FCI; [20]), the Mechanics Baseline Test (MBT; [21]), the Force and Motion Concept Evaluation (FMCE; [22]), the Maryland Physics Expectations Survey (MPEX; [23]), and the Conceptual Survey of Electricity and Magnetism (CSEM; [24]), did not mention similarities or differences in the performance of men and women in the articles that introduced the instrument.

Since its original publication, the FCI has perhaps been studied more than any other conceptual test. There are several articles about achievement gaps in the FCI, and one article about the creation of a more gender-fair FCI discussed below. Jennifer Docktor and Kenneth Heller published FCI scores for more than 5000 students, of whom 20% are women, taken over a 10-year period. They found a consistent gap averaging over 15% in pretest scores, and another consistent gap of greater than 13% in post-test scores. Both men and women made gains representing approximately six questions on the FCI during the term [25]. Furthermore, they found a stronger correlation between post-test scores and course grades for men than for women; the gender difference in course grades was not significant. One way to interpret these findings is that FCI post-tests correlate with course grades for women but over-predict course grades for men. At both a different university and high school, Coletta and colleagues found gender gaps on normalized gains on the FCI despite no gaps in course grades [26]. These gaps were largest for students with high scores on the FCI and Lawson Classroom Test of Scientific Reasoning, but were reversed for students with the lowest Lawson scores [26].

Another study challenged the FCI, noting that several of its questions included contexts that were stereotypically masculine (hockey, cannonballs, and rockets). A Revised Force

Concept Inventory (RFCI) was written, keeping the physics of each item exactly the same but changing three things: all figures depicted were of females, the context of the questions became stereotypically feminine (shopping, cooking, and jewelry), and abstract classroom lab situations were changed to be about daily life [27]. The RFCI was piloted by college students in English, sociology, and math classes; some students took the original FCI and some the RFCI. The gender gap on the RFCI was smaller than on the FCI. This was not, however, because women performed better on the RFCI than on the FCI. Instead, men did significantly worse on the RFCI than on the FCI [27]. Thus, the context of our assessments matters.

Sometimes the developers of standardized measures have addressed gender differences in their articles introducing the measures. The first was an article regarding the Test of Understanding Graphics in Kinematics (TUG-K; [28]). More recently, developers of the Determining and Interpreting Resistive Electric Circuit Concepts Tests (DIRECT; [29]) and the Colorado Learning Attitudes about Science Survey (CLASS; [30]) discussed gender differences found in the development of their measures.

When Engelhardt and Beichner developed the DIRECT, they noted that girls and women did not perform as well on the test as boys and men, either in high school or at universities. Furthermore, interviews showed that, at universities, women had more misconceptions about DC circuits than men. However, this difference was not observed between boys and girls in high school. Moreover, male students were much more confident in their answers than female students, though reasons for this discrepancy were not explored [29].

The developers of the CLASS also looked at gender differences [30]. Their instrument has 42 statements that students can agree or disagree with, and the responses of men and women differ significantly for more than half of the statements. In addition, women students' responses are less expert-like than those of men for several categories of statements. Women students at all levels are less expert-like in their views about real-world connections to science and their personal interest in physics [30]. The developers of the CLASS define "expert"-like views as views consistent with those of physics faculty; given the gender makeup of the faculty in our field, we wonder if expert views would naturally be more masculine views.

### B. Interest, Preparation, and Retention

Diekman and her group have found strong ties between women's interests and goals in pursuit of careers in science, technology, engineering and mathematics (STEM). They found that women have more communal goals than men; their focus is more on helping others than on helping themselves [31]. STEM careers, however, were not viewed as careers in which people help others. When students were given examples of STEM careers in which people do help others, their opinions could change (pages 910-911).

Hazari and colleagues noted that women students entered the university with higher grades than men, but received the same grades in their introductory physics classes as men; that is, they were underperforming [32]. Looking at dozens of variables relating to pre-college experiences, they found several things that helped both men and women. The first was strong academic backgrounds, measured by math SAT scores, high school calculus enrollment, and good grades in high school English, math, and science. There were also aspects of a high school physics class that helped both men and women: covering topics for longer periods of time, discussing the history of physics, incorporating videos into the classroom, and including test questions that involved calculations. Other things that helped women (more than men) were having a high school physics course that emphasized understanding over memorization and having a father who encouraged them to take science classes. Further work by the same group [33] surveyed only high-school girls. They tested five hypotheses about what might affect girls' interest in a physical science career: single-gender physics classes, women scientists as guest speakers, women physics teachers, discussing the work of women scientists during class, and discussing the underrepresentation of women in physics. Of these variables, only the discussion of the underrepresentation of women positively correlated with an increase in girls' interest in a career in the physical sciences [33].

Seymour [34] found that women leave STEM fields at a higher rate than men, even though their preparation and GPAs are just as good as those of men. Women also have different complaints about the teaching of STEM classes compared to men. Whereas men complained about large introductory classes having too much competition and being taught by less-qualified faculty, women noted microaggressions from the faculty and found it difficult to learn from professors who did not take a personal interest in them. In fact, the women students had more complaints about traditional university STEM classes than men. Women students complained about the impersonal nature of their classes and noted that their professors did not appear to care about them or even know them [34].

### C. The Effects of Reformed Pedagogy

Since Hake published his oft-cited paper in the American Journal of Physics [7], the PER community has agreed that interactive-engagement, or reformed, pedagogy is superior to traditional "transmissionist" methods of teaching physics. Breaking down achievement by gender (measured by gain scores on the FCI) was beyond the scope of Hake's study. However, several more recent papers about certain reformed pedagogies have discussed their effects on women students in particular.

Student-Centered Activities for Large-Enrollment Undergraduate Program (SCALE-UP) classes, started at North Carolina State University, have been shown in particular to help women and minority students [35]. Similarly, the Investigative Science Learning Environment (ISLE) laboratories, developed at Rutgers University [36], help women students make up for initial deficits and pass their first-year physics course at the same rate as the entire class [37].

At Rutgers, an extended, higher-credit-hour introductory course was developed for students who entered the university with low math skills. This course incorporates more active learning, and is where the ISLE labs were developed [36]. By the time this new course had been in place for a few years, women who had taken the course were just as likely as men to both pass the course and finish STEM degrees. Neither of those rates had been similar just a few years before [37].

A group at Harvard University reported eliminating the gender gap on the FCI by using Peer Instruction [38], the University of Washington Tutorials [39], and cooperative group problem solving [15,40]. They emphasized that closing the gap was due to women improving their performance while men continued to perform at high levels [41]. However, this result could not be replicated at the University of Colorado [42].

The group at Colorado was able to reduce the gender gap in course grades and FMCE scores with a brief values-affirmation exercise [43]. This was not, however, easy to replicate at the same institution [44].

### D. Questioning Standards of Measure

There are two issues with nearly all of this work. One is that it has taken an uncritical look at sex and gender as binary categories, when the reality is much more complicated. The other issue is that when a gap is noted between men and women, it is generally framed either implicitly or explicitly as "why can't women be more like men?" No one questions whether getting higher scores on standardized measures and persisting through physics majors is a good idea for both men and women. Is the goal to change women so that they can succeed in a culture where men are successful, or would it be better to change the culture so that the experience of men, particularly straight, white, married men, is not assumed to be the best standard?

The culture of physics, including physics departments and physics courses, has been designed by and for men [45][3]. The idea of physics, and science itself, as a male domain is persistent and can keep women from being interested in science [46,47]. One of the fathers of modern science, Francis Bacon, also famously talked about the earth and nature as female. Speaking of the practice of science, he used metaphors of marriage and rape as he talked about how men could discover the hidden secrets of nature [48]. Whether one places much significance in the metaphors used by a philosopher of science four centuries ago, it remains true that science has been practiced by men more often than women. Perhaps because of this imbalance, the qualities of good scientists have come to be associated with men, in the same way that the qualities of good nurturers have come to be associated with women. Some feminist theorists have suggested that the questions, practices, and answers of science might be different if scientists and decision makers were women rather than men [45,49]. Currently, many stereotypes abound in Western technological culture that relate to both science and sex differences; good scientists, and good men, are knowers, rational, and predictable. Women are framed as emotional, unpredictable, and thus irrational and poorly suited to science.

---

[3] Some readers (regardless of gender) may find this statement objectionable, and others consider it obvious. For those in the former group, we highly recommend Schiebinger's work [45], which develops the argument from both a historical perspective on the discipline and its modern-day form.

Furthermore, much of the PER work cited above has used standardized measures to compare men to women. As noted, most of these measures were developed without validations to determine if they were fair or equitable. Ignoring women as research subjects is a long-documented problem in medicine, paleontology, and the biological sciences, and remedying this deficiency in research data has benefited these fields [45]. Until balancing measures were taken, it was common for women to not be recruited as study participants, for assumed gender roles to be projected on them without verification, or for them to be treated as occasional deviants from the male norm [45]. An argument can be made that similar efforts are needed in PER: a recent review of gender gaps on conceptual diagnostics noted that "Since average normalized [FCI] gains are larger for men than for women, it is possible that having more women in a class could reduce the overall normalized gain for the class, thus making a teaching method appear to be less effective than it might appear in a class with a larger proportion of men" ( [50], p. 13). Although it was likely not the authors' intention to suggest that women in physics classes are damaging to the cause of education reform, the framing is telling as to which group is considered the problem. The paper contains extensive and careful discussion of remedying women's possible deficiencies on the tests, but no comment on the culture of physics as a factor that might contribute to or reinforce the reported gender gaps.

A much smaller but very useful body of research refrains from comparing women to men, but instead studies and celebrates differences among women. Some women who choose to study physics do it because they want to be useful. However, some women really like math and figuring things out. This work is highlighted below, but first we discuss one possible alternative to the binary deficit-based model.

### III. GENDER PERFORMATIVITY

As outlined above, research concerning gender within the PER community has largely dichotomized gender as a binary system of "man" and "woman". The absence of a more detailed discussion of gender within this body of literature suggests an underlying epistemology so strong and solidified that it goes unnoticed or unmentioned. This implicit framework limits the research questions we can ask, in the same way that focusing exclusively on student misconceptions constrains awareness of students' productive resources for learning [9]. An important step forward for PER is to embed a theoretical understanding of gender in our research questions, chosen subjects of study, and evaluation of evidence. This understanding could better support progressive and transformative research, which has long been a strength of the PER field.

Gender performativity is one such theory that has had important impacts on the larger academic community's understanding of gender. This theory has strong underpinnings and would help further refine the current efforts of many physics education researchers to support gender diversity in the classroom. We first discuss performance theory as a framework, then give examples of its application in several areas including in physics education.

## A. Description of the model

The underlying assumption of a clean divide between genders is that of biological determinism. Specifically, it is considered a "given" that nature produced a system of two genders that dictates many social realities that have real-world consequences (for example, the longstanding division of home labor as "women's work"; see [49] for one discussion). Butler's performance theory confronts these biological assumptions and agrees with other work that discusses how gender is in fact a social construction that we create through our performance. Butler suggests that "Gender reality is performative which means, quite simply, that it is real only to the extent that it is performed" ( [52], p. 527). By performance, Butler means that gender is something that is done—it is enacted, rather than a predetermined state. Gender is performed through dress, speech, how a person composes themselves, the jobs and passions they pursue, and more [53].

In the same way that a Shakespearean actor may channel Macbeth or Romeo but is not actually those characters, a person channels their gender [52]. It is an act, both conscious and unconscious. The script of this act is given to them through repetitive social interactions with parents, siblings, peers, teachers, movies, books, music, and beyond. By watching, reading about, and listening to others, an individual builds a social construct of how someone of their particular gender should act. An individual then rehearses these acts through their own actions (or inactions). The continuation of these gendered tropes by subsequent generations is what allows a lasting historical idea of gender to survive. Butler explains:

> "gender is an act which has been rehearsed, much as a script survives the particular actors who make use of it, but which requires individual actors in order to be actualized and reproduced as reality once again" (526).

In this light, gender is defined by social circumstance and is not a fixed reality. It is an illusion that has to be continually reinforced by social performance. A man who wears pants and acts tough, competitive, and in a manner to be viewed as a leader (all stereotypically masculine traits) would most likely not come to his workplace in a dress. To do so would perform the gender opposite of what he intended and what his social interactions have told him to perform. The social performance of wearing a dress would trigger the communication of woman, which socially is not defined by the characteristics the man was trying to embody, such as toughness and competition [54]. Gender is an ongoing performance:

> "gender is in no way a stable identity or locus of agency from which various acts proceed; rather, it is an identity tenuously constituted in time—an identity instituted through stylized repetition of acts. Further, gender is instituted through the stylization of the body and, hence, must be understood as the mundane way in which bodily gestures, movements, and enactments of various kinds constitute the illusion of an abiding gendered self" ( [52], 519).

Ideas of gender can change from one culture to another, and may not be constant between different societies. Although the man discussed above would most likely not wear a dress in North American culture, an otherwise similar man in Scotland might safely wear a kilt in order to represent his gender in his social context. In the same way that gender performances can change from country to country, they also change over time. This is easily seen when considering early Hollywood movies. Before the 1930s, both socially and in the movies, women were expected to wear dresses. These gendered assumptions of dress and ability to perform "woman" were challenged by actress Katharine Hepburn in her early films where she wore pants, which was considered controversial at the time. Ms. Hepburn once explained in an interview: "I wore pants when they weren't fashionable; I sat down on the curb if I was tired; I did what I wanted and what I thought was reasonable so long as I didn't hurt anyone" [55]. Today, women wearing pants in a movie is commonplace, and not considered material for tabloid gossip or concern in an interview. In the above examples and in many other ways, gender performance can change over time and space.

The social tolerance of varying genders is particularly evident when one looks to native inhabitants of the Americas. Native Americans are well known for their broader understanding of gender that was acted out in their society before the invasion, and impending genocide, by European travelers [56]. In some tribes, men and women were allowed to "change" their genders and take on roles traditionally associated with the gender that was not assigned to them at birth. These persons were known as "Two-Spirits" because they were believed to embody both masculine and feminine characteristics [57]. For example, a child assigned male at birth could choose to dress as a woman, partake in women's activities, and be married to men [57]. Children assigned female at birth were afforded the same options as well. One particular indigenous group in Mexico, the Zapotec, also allowed gender flexibility for their male sons [58]. Sons in this culture were allowed to dress as women and date men. This group, or third gender, is known as the Muxe. Muxes were embraced by the Zapotec community and played an important role in the family[4]. They were often the caregivers of aging parents as their siblings would pair off into heterosexual marriages and leave the home [59]. This role of supporting heterosexual marriage is different from the Two-Spirits, who were valued for the strength of possessing both male and female qualities and often became spiritual leaders. When we look at these examples, we can see that how gender is enacted and the rules governing allowed performances can extend beyond a simple binary but are always embedded in a community. Indeed, some researchers examine masculinity and femininity as localized communities of practice in their own right [60].

---

[4] Even when looking at the example of the Muxe, we see that gender variability and acceptance is still used as a method to promote heterosexuality. Although the Zapotec allowed a third gender, it was for the purpose of child rearing: the Muxe looked after aging heterosexual parents so their heterosexual siblings could leave the home and pursue their own families. When childbirth mortality was high for mothers and infants, creating stable gendered roles was deemed a necessity to ensure the creation of more humans. In the modern era, these concerns have largely been mitigated, at least in settings where physics education researchers work. Even absent this survival-based push, in our culture, gender is still thoroughly enforced in what people wear and how they act. It thus remains a way to control people in body and social access. It guides women and men toward, and away from, certain careers while also reifying men's overall dominance in the workplace, government, and home [49]. It provides a reason for differences, one that may advantage certain groups.

The historical and social nature of gender makes the mere questioning of its existence challenging, uncommon, and something that can trigger confrontation and anger. Gendered expectations in society are so ingrained and pervasive that many individuals feel they decide to present themselves and act in certain stereotypically-gendered ways entirely out of choice. However, Butler would argue that these individuals think they are objective, when in reality their perceived objectivity was constructed from the expectations surrounding them [51,52]. With the pervasive nature of gender socialization in mind, it is understandable that PER itself has never challenged its underlying gender epistemologies.

Within this paradigm, Butler also examined the idea of sex, or rather our biological gender. Butler argues that sex itself is also a social construct because the roles associated with sex are predicated on the performance of gender. This argument goes beyond the scope of the present article, but is important to recognize. Beyond Butler's [51,52] philosophical musings, the theory of performance has been used in sociological work, including a few studies tackling issues within the realm of physics.

### B. Applications of performativity

Gender performance has been analyzed in many qualitative projects in the field of education. Such studies have examined the construction of gender for male elementary teachers, lesbian women in sports, men in undergraduate institutions, and women in physics  [61–64].

Gender roles encompass the jobs we pursue and often dictate fields that become dominated by one gender. One such field is elementary education, where women are the majority. Recent calls have even asserted the importance of male teachers, in essence to teach masculinity and relate to young male students. One study examined the breadth of ways that three male teachers performed their genders, and argued that it is flawed to assume the presence of men will teach any single unified gender [61]. A similar study on masculine performance considered the disconnect between males' explained understanding of gender and their actual performance of gender [65]. Findings indicated that male undergraduate students espoused complicated understandings of gender that allowed men to act in many ways, but actually performed in stereotypically masculine ways (as identified by the authors of that study), such as being misogynistic, consuming copious amounts of alcohol, and acting homophobic to male peers [65].

Studies have also aimed to understand issues of femininity as well. Powell and collaborators [66] explored women's performance of gender in the field of engineering. In this setting, the authors found women who often transformed into "honorary men". Women did this through a gendered performance that included acting like "one of the boys," achieving a high reputation, and becoming anti-woman themselves. Through these efforts, these women worked to be seen as men and not as women. Tonso [67] studied students at an engineering school and found that the range of identity labels available to women was much smaller than for men. Women's identities (as assigned by other students) centered around campus social roles or a few academic characteristics such as "hard-worker," but none of the more positive technically competent "Nerd" category of labels. Pressure to disregard femininity in order to be included in the field has also been observed in women's performance of gender in physics.

Qualitative studies have found that some women in graduate physics programs actually identify feminine characteristics as contradicting logic and physics [63]. In this same study, the culture of physics was described as being inherently gender-neutral but one where masculine traits dominate and feminine traits were viewed as negative. For example, feminine dress (such as wearing dresses and heeled shoes) was explained by the participants as something that would look out of place in the physics lab. One participant explained that she was told not to dress in this way so that supervisors would always think she was working. Makeup, dresses, and heels were said to demonstrate a focus on one's self, and in physics all attention must be paid to one's research [63].

Gonsalves explained in another piece that "…the symbolic masculinity of physics reifies an understanding of women as an always, already gendered category that is naturally situated in opposition to physics ( [68], p. 119). Gonsalves' work explores the tension that arises when physics is viewed as inherently masculine, which in turn threatens women's participation because they are immediately seen in contrast to the physics culture. What is particularly alarming is that this masculine culture is painted as gender-neutral, thus obscuring and normalizing gendered expectations in physics.[5] The women in Gonsalves' study [63] demonstrate this conclusion through their rejection of femininity and efforts to become masculine. Similar conclusions were also reached in Sharon Traweek's seminal ethnographic account of high-energy physicists. Her work found a masculine culture described by its inhabitants as being objective and neutral, a "culture of no culture" [69].

These investigations demonstrate some of the ways that the study of gender through performativity is both illuminating and productive for work in PER. Danielsson [64] and Gonsalves [63] have charted the beginning of such explorations in the field, but many questions and discoveries await those who wish to take up the performativity lens.

## IV. GENDER, RACE, AND COMPLEXITIES

As highlighted in the previous section, gender performance is complex. This understanding can be expanded more broadly to discuss the complexity of individuals as a whole. When we think about the stereotypes of what makes a scientist, even an internet image search presents a stereotypical person (albeit in cartoon form). The image of a white male with crazy hair, wearing a lab coat, is so ingrained in society that children of many different ages typically recreate this image [70,71]. An image search for "physicist" returns primarily male dominant images. What effect, if any, does this embedded image of the white male scientist have on the choices for individuals who do not fit this norm?

---

[5] The finding of Hazari et al. [33] that explicit discussion of underrepresentation had a positive effect for women may connect with this point. Acknowledging the differential cultures faced by men and women in physics relieves women students, at least in the context of the discussion, of the additional invisible burden of pretending that the status quo is genderless.

Steinke and colleagues [72] suggest that the media perpetuates students' perceptions about scientists, and that short-term interventions are not sufficient to change these stereotypes.

Although we extensively discuss performance theory in this paper, other literature uses the term "identity". Identity includes—but is not limited to—race, ethnicity, gender, physical ability, sexual orientation, religion, socioeconomic status, and political affiliation. In this section, we consider many aspects of identity as a performance, but use similar terminology presented in the published literature. Certain identities are easily performed and made visible, others can be hidden, and some are blurred. For example, sexual orientation, religion, and political affiliation, if not discussed, are not known. Therefore, the performance is hidden or not displayed. Race, ethnicity, and some areas of physical ability are all visible markers of identity and typically cannot be hidden. Socio-economic status can be a visible marker (for example, wearing an expensive name brand of clothing) but can also be hidden. The most visible markers are typically the pieces of identity performance that distinguish inclusion or exclusion from a group [73]. Although identity, or identities, of an individual may be considered nothing more than performances acted out because of social constructs, these social constructs carry substantial real-world consequences. Microaggressions, macroaggressions, and stereotype threat are measurable consequences for populations that fall outside the categories of white, middle-class, heterosexual, non-disabled, and male. These real-world consequences are not built overnight but are accumulated over lifetimes and perpetuated based on what is supposed to be the "norm." Stereotype threat and microaggressions are typically attributed to these visible performances of race, gender, and so on.

Stereotype threat is defined as a situational predicament in which people are, or feel themselves to be, at risk of confirming negative stereotypes about their social group. Stereotype threat is linked to a number of negative academic outcomes and is stronger when the threatened identity is primed before or during a difficult task. For example, African Americans who are asked to write down their race before taking a standardized assessment typically do poorer on those assessments than a control group of students who were not asked to write their ethnicity until the exam was over [74]. The negative stereotype for this example is that African Americans are not smart. Another example of relevance of physics classes examines stereotypes between women and math ability. When women are reminded that they are not expected to be as skilled at math as their male counterparts, their scores drop [75].

Stereotype threats are self-fulfilling by the individual. Microaggressions, in contrast, come from peers or superiors. Microaggressions are brief and common daily verbal, behavioral, or environmental indignities, whether intentional or unintentional, that communicate hostile, derogatory, or negative slights and insults [76–78]. Maria Ong [79] explicitly studied women of color in physics as a case for the intersection of race, gender, and science. She argued that women of color often employ fragmentation strategies, which include gendered passing and racial passing (i.e., actively or passively seeking to be perceived and accepted as a member of a more dominant group). In Ong's work, subjects' use of this strategy serves "to achieve one of two performance-related goals: 1.

to organize themselves to be seen as community members or non-members, or 2. to organize the appearance of competence" (p. 595). She describes several instances of women of color of changing their femininity to be more male-like when working with their peers: a Chicana changed from a skirt to pants to go to an all-male lab, an African American woman spoke loudly when asking questions because males are loud and aggressive and that characteristic is more male, or changing the word choice from "I think" to "I know." These are examples of gender performance as described in the previous section. However, they are also "survival mechanisms" that occur in response to microaggressions in the physics environment.

Performance also played a role in Ong's study when discussing the ethnic identity of one of her female subjects. Her mother was born in Korea and is the child of a Korean woman and an American G.I. Her father is half Black and half Native American, so she is "very mixed". In her courses, however, she took on the role of the "loud black woman" in order to become hypervisible. Again, this is a survival mechanism of learning how to fit in by not fitting in. In other words, in order to survive, women of color learn how to blend in better or learn how to be extreme enough to stand out.

Similarly, one author of this paper has described similar feelings of learning how to blend in. She is Chicana and Native American. She has been the only female, as well as the only ethnic minority, in some of her courses and has dealt with a multitude of micro- and macroaggressions at all levels of her career. She described changes to her female identity from more masculine (to blend in) to more feminine (to become hypervisible) until she found a level that feels natural and unperformed. The progression of the performance of her identity has grown from learning to be confident without having to match her peers. Her cultural performance, however, has not gone through similar levels of blending in. She has described her cultural identity as always being hypervisible because she is immensely proud of her *ChicanIndia* (the author's term) identity, which is always apparent. She displays her cultural heritage in her style of attire, her earrings, or other accessories. Her performance as a scientist is the least defined, has gone through the most levels of blending. It remains less defined depending on the audience with which she is engaged. Her *ChicanIndia* self is easy for her to define because she has grown up with cultural mentors. Her physicist self is still developing because she has never known a physicist culturally like her.

Li and Loverude found that upper division chemistry and physics students from diverse backgrounds also change their performance depending on the community in which they are engaged [80]. When talking with individuals who are not scientists, some students refer to physics or chemistry as a negative experience because they want to relate to the general perception that science is hard and not fun. The same students, when talking with other science students, will discuss physics or chemistry as a positive experience because they are around an audience that has a similar, positive response to these experiences. Therefore, the negative or positive performance changes depending on the audience. These opposing performances create conflicts when trying to find a niche as a physicist in physics culture, described by Traweek as a "culture of no culture" [69].

Similar themes of hypervisibility, clashing identities, and gender performance are explored by Faulkner [81,82]. Her study of engineering workplace cultures explores the limited and gender-marked range of engineering identities available, as well as what she calls the 'in/visibility paradox" faced by women who are simultaneously less visible as engineers while always perceived as women. Faulkner [82] discusses

> "If to be a 'real engineer' is to be a man, and if 'men' and 'women' are necessarily different, then women engineers have to play down their identity as 'real women' if they are to belong in engineering. Although women engineers are highly visible as women, they must also learn to, in some sense, become invisible as women" (179).

This conflict is even more pronounced for women of color in STEM fields. The performance of different identities creates higher levels of stress for individuals who belong to more than one group that is underrepresented in physics. For example in a study by Ko et al. [83] examining narratives of women of color in physics, astrophysics, and astronomy, one participant says:

> "[Race and gender] aren't separate in me. I am always black and female. I can't say, 'Well, that was just a sexist remark' without wondering would he have made the same sexist remark to a white woman. So, does that make it a racist, sexist remark? You know, I don't know. And that takes a lot of energy to be constantly trying to figure out which one it is...somebody has some issues about me...being black, female, and wanting to do science and be taken seriously" (222).

It is known that women and ethnic minorities have low representation in STEM fields (see note 1), particularly in physics. According to data collected by the American Physical Society in 2012, both engineering and physics awarded 20% of their bachelor's degrees to women. If we look at APS data from 2012 for Hispanics, African Americans, and Native Americans, the numbers are as follows: 8% for bachelor's degrees, 5.5% for master's degrees, and 4% for doctorates in physics. These numbers outline the greater issue of inclusion or exclusion in physics. Miller and Stassun [84] have highlighted GRE admission scores as an important area of exclusion. They describe the GRE as "a better indicator of sex and skin color than of ability and ultimate success." A common cutoff score for physical science Ph.D. programs for the mathematics portion of the GRE is 700. According to their study, "only 26% of women, compared with 73% of men, score above 700 on the GRE Quantitative measure. For ethnic minorities, this falls to 5.2%, compared with 82% for whites and Asians." This has huge implications about who represents the included majority. This work and other research that focuses on self-efficacy [85] describes fitting into the physics culture as it currently stands, but it also raises the question previously expressed: Is the goal to change women so that they can succeed in a culture where men are successful, or would it be better to change the culture so that the experience of (straight white married male) men is not the assumed standard?

Gender performance is much more complex than the present PER literature describes, but even complexities within gender limit our understanding of ethnic minority populations. When we go a step further and examine the performance of multiple identities, our understanding becomes even more problematic. The desire to generalize results across studies causes comparisons of how "this" population relates to the "standard" population, which implicitly declares a standard that is seldom carefully examined. The standard then reinforces that students should be more like the "norm" (white males) and when "this" population shows a difference, it is implicitly suggested that the "standard" is better. As researchers, we simplify our analyses, particularly in the quantitative domain, by assuming that student identities fall into simple, discrete categories of gender, race, and so on. But this process of simplification often obscures the very details of learning that it seeks to uncover—and worse, does injustice to students who in one or in many ways do not identify with the "culture of no culture."

## V. MOVING FORWARD: SUGGESTIONS FOR FUTURE WORK

…the feminist empiricist strategy argues that sexism and androcentrism are social biases, prejudices based on false beliefs . . . and on hostile attitudes. These prejudices enter research particularly at the stage of the identification and definition of scientific problems, but also in the design of research and in the collection and interpretation of evidence. According to this strategy, such biases can be eliminated by stricter adherence to the existing norms of scientific inquiry…  [49]

In the preceding sections, we have reviewed work on gender and other identity performances, and on gender specifically within PER. We see that most of this work lies in a (usually unarticulated) framework of gender as a strict binary, often conflated with sex. Among several damaging consequences of this framework is that it invites a deficit model, wherein female students are presented as lacking some combination of science-like traits (math preparation, or self-confidence). Thus, the implied solution is to help women be more like men. However, there are deeper structural issues with this model that caution against its use as the sole scaffold for research on gender in physics education. In particular, the constraint of two genders restricts student identities for the purposes of designing research questions, collecting data, and reaching conclusions. It also ignores the intersection of gender with race or ethnicity, socioeconomic status, LGBTQ identity, and other aspects of identity that students perform in the varied contexts of their lives.

Going forward, we offer a number of possible suggestions for research directions that expand beyond the gender binary deficit model. This list is by no means exhaustive; it is intended to provoke discussion and offer options to those who are troubled by the above theoretical considerations and interested in concrete applications. For organizational purposes, we divide the recommendations into three categories: theoretical frameworks, methodology, and subjects of study. Though many of the individual points share some overlap, they each highlight different examples and themes for development.

### A. Recommendations around theoretical frameworks

The debate continues among physics education researchers over the importance of using and articulating theoretical frameworks. One argument in favor says that when we fail to be explicit, we do not actually avoid having and using theories. We only default to less specific and less examined beliefs under the assumption that everybody shares our foundations for deciding worthwhile topics of study, specifying research questions, and evaluating evidence as compelling or insufficient. Additionally, even the same set of data can yield very different insights depending on the theoretical lens employed to examine it [86,87]. Accordingly, we suggest some large-scale perspective shifts that might occur.

### 1. Projects that transcend a "gender gap" framework

One of the most straightforward ways to avoid problematizing differences between students is to step away from "gap-gazing." Focusing on achievement gaps to the exclusion of other considerations reinforces deficit models, implicitly positions tests (and the culture that produced them) as unbiased, and often reduces student identities to essentialized categories (e.g., "all women…"). Studies that examine student identities with an explicitly anti-essentializing methodology can yield very different insights. Examples include Gwenyth Hughes' study of science identity construction that intentionally avoids pre-classifying students by gender binary expectations [88], or work by Karen Tonso [67,89] and Wendy Faulkner [81,82] in engineering that explores a range of gender performances by women and men. For researchers interested in exploring a more fluid range of gender identities in physics, the methods and findings of these papers are one place to start.

Other studies explore multiple facets of student participation and success in physics [64,90–92], or compare between different groups of women [91,93,94]. Any of these approaches can give a more nuanced picture than reducing students to clearly-marked binaries and placing them in opposition. One more suggestion the authors can make in this area is to be more explicit about the demographics of the populations studied. The work need not be comparative in nature, but clearly describing the demographics gives more context about how the study can be applied to other populations.

### 2. Explicitly feminist projects

One key to understanding the complexity of identity has been a shift in researchers' perspectives from observer to actor. As an observer, an "impartial" researcher assigns research variables of interest according to their own preconceptions, which may or may not be articulated in a theoretical framework. As an actor, gaining perspective on the lived experiences of subjects is used to build a more authentic picture of the situational factors that are most relevant to their lives. Arguably, one of the disciplinary foundations of PER is the shift away from a strict faculty-centered (observer) perspective to consider the viewpoints and knowledge of student actors.

The choice of research framework by an "objective" outside researcher has been discussed at great length in the context of gender in science. Much of this work has occurred in women's studies or gender studies, but key outcomes have been summarized in non-specialist sources (see [45] for one introduction). Roychoudhury, Tippins, and Nichols [95] provide one example of a classroom study using feminist standpoint theory as its theoretical framework. Bug [11] outlines some of the interplay between physics education reform and equity projects of including more women in physics. In particular, although low representation of women is recognized as a problem to be remedied, such equity-by numbers projects often neglect considerations that the structures of academic science favor men. Instead, the focus is largely on "physics education as usual," but in a fashion that will produce higher gains on better tests. This tight focus on one theoretical framework for gender neglects the interdisciplinary strengths of PER. Whitten [96] outlines nine categories of potentially feminist physics projects. One of these explicitly deals with changing physics education, but others include "projects that problematize the knowing subject/object of inquiry split" and "projects that reconceptualize physics in less reductionist directions," labels which could also describe many PER studies. Thus, we suggest that physics education researchers would find deeper exploration into feminist critiques of science to be complementary to their interests in many cases. If nothing else, feminist critiques of problematics (e.g., who benefits from this research? what constitutes "objectivity?") are of great value to researchers whose work is intertwined with human subject considerations.

### 3. Studies using feminist-friendly theoretical frameworks

Brickhouse [97] describes how sources on feminist epistemology exist in much greater quantities than those on feminist pedagogy, which in turn is much more explored than feminist theories of learning. She proposes that constructivism has been taken up somewhat opportunistically by feminist education researchers and argues that a more purposeful selection of research frameworks can be made. Brickhouse suggests situated cognition [98,99] as a theory of learning that shares substantial overlap with feminist epistemologies, and incorporates a focus on identity development. A number of physics education researchers have already made use of the literature on situated cognition and communities of practice. Thus, an expanded treatment of gender is one logical extension of these investigations. Preliminary work by Paechter [60] has outlined a case for masculinities and femininities as communities of practice as one promising avenue of development. Faulkner [81,82] develops this theme by studying workplace communities of practice in engineering and the types of masculinities and femininities that are available (or not available) to members.

### B. Recommendations around methodology

Even after the field of interesting research questions has been somewhat constrained by a theoretical framework, individual investigations and programs of study encompass a variety of methodological choices. Here we highlight two possibilities that are aligned with the arguments in this article.

### 1. More qualitative work on gender

A formal background in physics, as well as residence in physics departments, understandably biases many physics education researchers toward quantitative study. However, Danielsson's recent review [100] points to a serious lack of qualitative studies on gender in physics education. This shortage is exacerbated when searching for intersectional work that incorporates other facets of identity [101]. These reviews highlight literature gaps in more detail and provide many suggested avenues. Qualitative approaches include free responses, interviews, analysis of videotaped course discussions, etc. as defined in the literature base [102,103], and explore the "actor" perspective of students in rich detail. These avenues open a range of critical questions even to "solo" physics education researchers at small institutions.

### 2. Quantitative work that attends to the complexity of identity

We do not mean to suggest that it is impossible to conduct quantitative studies that acknowledge the multifaceted nature of identity. One example of such an investigation is work done on stereotype threat, which has grown from an initial investigation of racial differences on a difficult verbal test [74] to a complex model of "identity contingencies" [104] that each person carries because of their gender, race, age, economic status, and other aspects of identity. Identity contingencies can depend highly on the situation: white male engineering students, typically not subject to stereotype threat in mathematics, showed a sudden drop in scores when told that their performance was being used to help understand Asian students' superiority in the subject [105]. Although some research has investigated the use of stereotype threat interventions in physics classes [43,44], much more could be done to probe the effect of identity contingencies on these student populations. Such work could also focus on faculty identities, such as climate research and studies of workplace experience. An example of a quantitative study looking at LGBT STEM faculty was conducted by Patridge and collaborators [94].

### C. Recommendations around subjects of study

Physics education research potentially focuses on a wide spectrum of "units of analysis:" individual student or faculty cases, average class scores on assorted diagnostic measures, department-wide teaching practices, large-scale surveys of students at many universities, and even international comparisons. Here we consider two candidates for this choice to deepen our understanding of gender in physics education.

### 1. Closer attention to group dynamics

Although peer interactions play a role (often a major one) in most curricula produced by the physics education research community, student group interactions are peripheral to or absent from many reports of results. This omission becomes more serious when considering gender—especially when some of the few published results for undergraduate physics indicate that gendered group dynamics can directly affect women's participation [15]. Most research on these specific issues has been conducted at the secondary level [106,107]. Researchers concerned with undergraduate settings may find useful references in the literature on engineering education, where teamwork has received more extensive attention especially in the context of design. Tonso [89] and references therein provide one stepping-off place for both teamwork and gender.

### 2. Exploration of "what works" findings

Whitten and collaborators ( [108] for summary, [109–111] for details) present results of an exploration into the "best practices" of physics departments that graduate a large number of women with physics degrees. They outline factors such as supportive departmental climate, personal attention to majors in the first year, and offering of stigma-free extra preparation while maintaining high standards. In addition to providing suggestions for any department looking to become more inclusive, these and other "what works" diversity findings in STEM [112–114] highlight areas of interest that are ripe for further study. Here, researchers can learn more about the experiences of successful students who do not necessarily belong to the "standard" population. Findings that attend to the intersection of multiple identities are especially important [94,114], as they may promote modes for students that require less fragmentation or hypervisibility (e.g., the "honorary male" or the "loud black woman").

## VI. CONCLUSIONS

In this paper, we have presented the case that gender-based PER to date has tended to focus along relatively narrow lines of investigation. These lines were not consciously chosen in many cases but have been carved by an extensive and culturally inherited framework of gender roles and expectations. Binary gender models often contain deficit implications and constrain research; however, they remain popular, even in cases where these models do not encompass the researchers' personal beliefs about gender. It is difficult to abandon such a culturally embedded framework without a compelling alternative. As one example, we have presented the performativity theory elaborated by Judith Butler. In discussing gender as performance, both generally and in physics-specific applications, we illustrate new dimensions of questions and answers that are thereby opened to researchers. These dimensions multiply when other aspects of students' identities are considered; we have merely discussed the intersection of race and gender as one example. The performance qualities of gender are often most visible to people who are caught at these intersections, who often must choose to perform one identity facet more heavily than another depending on the community context of the moment. Instruments or studies that reduce this complexity of identity down to dichotomous check boxes should be treated with some caution and never be used as the final word on whether a particular instrument or curriculum is effective for "all students."

It is not our intent to argue that anyone studying gender in PER must first pursue a doctorate in women's studies, any more than physics education researchers coming from education or physics departments do secondary dissertations in the other discipline. However, we have referenced feminist critiques of science, examples of implicitly or explicitly feminist physics projects, and studies that examine the intersection of gender with other parts of identity performance. We hope that this range of examples will provide inspiration and some possible starting points for physics education researchers who wish to build on the foundation of gender in PER.

## ACKNOWLEDGEMENTS

R. Barthelemy was supported by a Fulbright Fellowship during this work. The authors also wish to thank the several colleagues who offered critical feedback and comments on the manuscript.